# The Algorithm Analysis of E-Commerce Security Issues for Online Payment Transaction System in Banking Technology

RAJU BARSKAR,
M.Tech. Student of CSE Dept.
MANIT, Bhopal (M.P) India
rajubarskar353@gmail.com

ANJANA JAYANT DEEN,
Faculty of CSE Department
UIT_RGPV, Bhopal (M.P) India

GULFISHAN FIRDOSE AHMED,
M.Tech. Student of CSE Dept.
LNCT, Bhopal (M.P) India
gul_firdose@rediffmail.com

JYOTI BHARTI,
Assistant Prof. IT Dept.
MANIT, Bhopal (M.P) India
jyoti2202@gmail.com

*Abstract:-E-Commerce offers the banking industry great opportunity, but also creates a set of new risks and vulnerability such as security threats. Information security, therefore, is an essential management and technical requirement for any efficient and effective Payment transaction activities over the internet. Still, its definition is a complex endeavor due to the constant technological and business change and requires a coordinated match of algorithm and technical solutions. E-commerce is not appropriate to all business transactions and, within e-commerce there is no one technology that can or should be appropriate to all requirements. E-commerce is not a new phenomenon; electronic markets, electronic data interchange and customer e-commerce. The use of electronic data interchanges as a universal and non-proprietary way of doing business. Through the electronic transaction the security is the most important phenomena to enhance the banking transaction security via payment transaction.*

*Categories and Subject Descriptions: -*
**Electronic commerce security, payment system, payment transaction security, payer anonymity, dual signature.**
*General terms:-*
**Electronic payment transaction security.**

## I. INTRODUCTION

An electronic payment transaction is an execution of a protocol by which amount of money is taken forms a payer and given to a payee. In a payment transaction we generally difference between the order information (goods or services to be paid for) and the payment instruction (e.g., credit card number). From a security perspective, these two pieces of information deserve special treatment. This paper describes some algorithm that can be used to implement the payment transaction security services.

## II. USER ANONYMITY AND LOCATION UNTRACEABILITY

User anonymity and location untraceability can be provided separately, a pure user anonymity security service would protect against disclosure of a user's identity. This can be achieved by, for example, a user's employing pseudonyms instead of his or her real name. However, if a network transaction can be traced back to the originating host, and if the host is used by a known user only, such type of anonymity is obviously not sufficient. A "pure" location untraceability security service would protect against disclosure of where a message originates. One possible solution is to route the network traffic through a set of "anonym zing" hosts, so that the traffic appears to originate form one of these hosts. However, this requires that at least one of the hosts on the network path be honest, if the traffic source is to remain truly anonymous.

### A. Chain of Mixes:-

A user anonymity and location untraceability mechanism based on a series of anonym zing hosts or mixes has been proposed by D. Chaum [1]. This mechanism, which is payment system independent, can also provide protection against traffic analysis. The basic idea is illustrated in Fig. [2.1] messages are sent from A,B and C (representing customer wishing to remain anonymous) to the mix, and from the mix to X,Y and Z )representing merchant or banks curious about the customer' identities). Messages are encrypted with the public key of the mix, $E_M$. if customer a wishes to send a message to merchant Y, A sends to the mix the following construct:

$$A \longrightarrow Mix:E_M[Mix, E_Y(Y, Message)]$$

Now the mix can decrypt it and send the result to Y:
$$Mix \longrightarrow Y: E_Y(Y, Message)$$

Only Y can read is since it is encrypted with Y's public key, EY. If the mix is honest, Y has no idea where the message originated or who sent it.
The main drawback of the scheme is that the mix has to be completely trustworthy.

If A wishes Y to send a reply, he can include an anonymous return address in the message to Y:
Mix, $E_M(A)$
In this way the reply message is actually sent to the mix, but only the mix knows whom to send it on to (i.e., who should ultimately receive it).

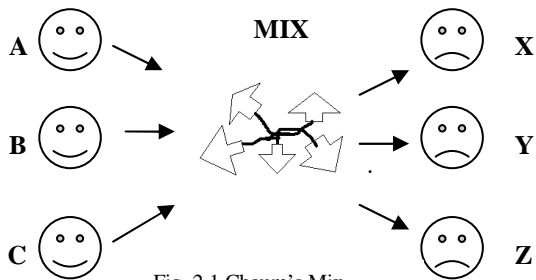

Fig. 2.1 Chaum's Mix

An additional property of the mix scheme is protection against traffic analysis.

This can be achieved by sending "dummy" message from A, B and C to the mix and from the mix to X, Y and Z. all messages, both dummy and genuine, must be random and of fixed length, and sent at a constant rate. Additionally, they must be broken into fixed block sizes and sent encrypted so that an eavesdropper cannot read them. The problem of having a mix trusted by all participant can be solved by using a matrix (or network) of mixes instead of just one, as shown in fig. [2.2]. in this case, only one mix on a randomly chosen path ("chain") has to be honest. The bigger the matrix, the higher the probability that there will be at least one honest mix on a randomly chosen path.

For a chain of mixes, let E, be the public key of Mix $i = 1, 2, 3$. A message is constructed recursively as follows:
E Recipient (Next recipient, E Next recipient (…))

If a wants to send an anonymous and untraceable message to Y, as in the example with on mix, the protocol goes as follows:

A     ⟶     Mix1:

$E_1(Mix_2, E_2(Mix_3, E_3, (Y, Message)))$

Mix1     ⟶     Mix2:$E_2(Mix3, E_3 (Y, Message))$

Mix2     ⟶     Mix3:$E_3 (Y, Message)$

Mix3     ⟶     Message

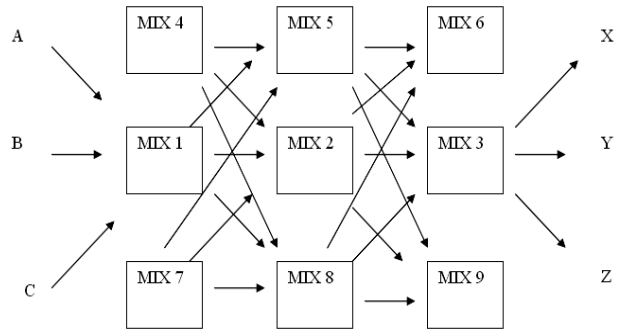

Fig-2.2 Chain of Mix

Message" can additionally be encrypted with Y's public key, which is omitted here for simplicity. The principal a can provide an anonymous return address in the same way as in the example with one mix. Specifically, A picks a random return path through the mix network (e.g., Mix2, Mix1) and encrypts his identity and address multiple times using the public keys of the mixes on the return path:
Mix2, $E_2(Mix1, E_1(A))$

The recipient of the message (Y) can then send the message to the first mix, and from this point on it works in the same way as from A to Y. An implementation of the mix network would be expensive and complex from both the technical and the organizational points of view. There is an experimental implementation of anonymous e-mail with return addresses called BABEl by Gulcu and Tsudik [3], and onion network.

## III. PAYER ANONYMITY
This is a simplest way to ensue payer anonymity with respect to the payee in for the payer to use pseudonyms instead of his or her real identity. If one wants to be sure that two different payment transactions by the same payer cannot be linked, then payment transaction untraceability must also be provided.

### A. Pseudonyms
The first internet payment system was based on the existing internet infrastructure that is e-mail, TELNET, S/MIME, and FINGER. Although they did not use cryptography at the beginning, they later realized that in some cases it was necessary. For example, authorization message exchanged between First Virtual and merchants before shipment must be protected to prevent large shipments to fraudulent customers. Under the First Virtual System, a customer obtains a Virtual PIN (VPIN), a string of alphanumeric which acts as a pseudonym for a credit card number. The Virtual PIN may be sent safety by e-mail. Even if it is stolen, an unauthorized customer cannot use it because all transactions are confirmed by e-mail before a credit card is charged. If some one tries to use a customer's Virtual PIN without authorization, First Virtual will be notified of the stolen Virtual PIN when the customer replies "Fraud" to First Victual's request for confirmation of the sale (Fig. 3.1) in such a case, the Virtual PIN will be canceled immediately. This mechanism also ensures confidentiality of payment instruction with request to the merchant and potential eavesdroppers. Fig. (3.1) illustrates a First Virtual (FV) payment transaction. A customer sends his order to the merchant together with his VPIN (1). The merchant may send VPIN authorization request to the FV payment provider (2). If the VPIN is valid (3), the merchant supplies the ordered services to the customer (4) and sends the transaction information to the FV provider (5). In the next step (6) the FV provider asks the customer whether he is willing to pay for the services (e.g., via e-mail). Note the customer may refuse to pay ("No") if the services were delivered but do not fulfill his expectations. If the services were not ordered by the customer, he responds with "Fraud". That aborts the transaction and revokes (i.e., declare invalid) the PIN. If the customer wants to pay, he responds with "Yes" (7). In this case the amount of sale is withdrawn from his account (8a) and deposited to the merchant's account (8b), involving a clearing transaction between the banks (9)

The payment transaction described above involves low risk if the services include information only. Even if a fraudulent customer does not pay for the services delivered, the merchant will not suffer a significant loss [4], and the VPIN will be blacklisted immediately, as mentioned before, cryptographically protected authorization message must be exchanged between First Virtual and merchant before large shipments.

## IV. PAYMENR TRANSACTION UNTRACEABILITY

Currently there is only one mechanism providing perfect anonymity and thus perfect payment untraceability. However, this mechanism (Blind signature) is used for digital coins.

### A. Randomized Has sum in SET
A merchant also obtains only the hash sum of a payment instruction. The payment instruction contains, among other information like: Primary account number, PAN (e.g., credit card number):The card's expiry date (Card Expiry);A secret value shared among the cardholder, the payment gateway, and the cardholder's certification authority (PAN Secret);A fresh nonce to prevent dictionary attacks (EXNonce).
Since the nonce is different for each payment transaction, the merchant cannot link two transactions even if the same PAN is used.

### B. Blind Signature
D.Chaum [1] proposed a cryptographic mechanism that can be used to blind (obscure) the connection between the coins issued and the identity of the customer who originally obtained them. The mechanism, which provides both payer anonymity and payment transaction untraceability, are based on the RSA signature and is called a blind signature. This type of signature is called blind since the signer cannot see what he signs. The basic scenario is the same as in RSA: $d$ is the signer's private key, e and n are the signer's public key. There is an additional parameter, k, called the blinding factor and chosen by the message (e.g., the digital money serial numbers) provider:

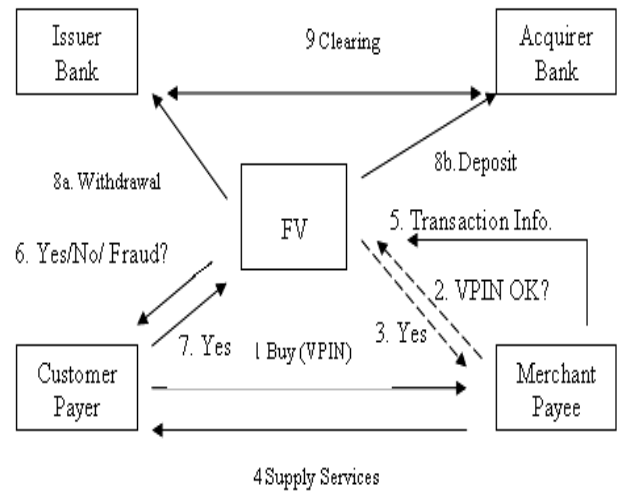

Fig. 3.1 First Virtual's Payment System

Provider blinds the massage M:
M = Mk$^e$ mod n;

Signer computers the blind signature:
S' = (M)$^d$ mod$n$ = kM$^d$ mod$n$;

Provider removes the blinding factor:
S = S'/k = M$^d$ mod$n$.

This signer usually wants to check if the message M(e.g., a vote or digital coin) in valid. For this purpose the provider prepares n messages and blinds each one with a different blinding factor. The signer then chosen n-1 messages at random and asks the provider to send the corresponding blinding factor. The signer checks the n-1 messages; if they are correct, he signs the remaining messages. Note that electronic coins blinded in this way can only be used in an online payment system.

## V. CONFIDENTIALITY OF PAYMENT TRANSACTION DATA

Payment transaction date generally consists of two parts: the payment instruction and the order information. A payment instruction can contain a credit card number or an account number. The primary purpose of protecting its confidentiality is to prevent misuse by unauthorized principal, including dishonest merchants. in many cases, however, the information contained in a payment instruction uniquely identifies the payer. Consequently, protecting it from unauthorized or dishonest principals also means protecting the payer's anonymity. Although a payment instruction and order information must some times be made unreadable to different parties, there must still be a connection between them that can be easily verified by the customer, the merchant and the payment gateway otherwise, in a case of dispute, the customer could not prove that the payment instruction he sent to the merchant really related to a particular order.

### A. Dual Signature

SET (secure electronic transaction) is an open specification for secure credit card transactions over open network. Its development was initiated by Visa and Master Card in 1996. SET uses the crypto technology of RSA Data Security, Inc., so it cannot be used without a license. There are other crypto libraries available that will work in the place of the default crypto library (BSAFE) To protect credit card number (or a customer's payment instructions in general) from both eavesdroppers and dishonest merchant, SET employs dual signature. In additional, dual signature protects confidentiality of purchase order information with respect to payment gateway. In a simplified scenario, let PI be the payment instruction and OI the order information. Let M be a merchant and P a payment gateway. We want the merchant M not to able to read the payment instruction PI, and the gateway P not to be able to read the order information OI. To achieve that, the customer computes the dual signature DS of the payment request. In other words, the customer C signs PI and OI intended for P and M, respectively, by applying a cryptographic hash function b() and his private key Dc from a public key algorithm:
Customer computes:

$$DS = DC (h(h(PI), h(OI)))$$

Since M is supposed to see OI only, and P to see PI only, they obtain the respective confidential part as a hash sum only:
Merchant Receives: OI, h(PI), DS
Payment Gateway Receives: PI, h(OI), DS
However, they can both verify the dual signature DS. If P agrees, that is, if the payment instruction in correct and the authorization response is positive, it can sign PI. If M agrees, he can sign OI. In the SET protocol the customer sends PI not to the gateway directly, but to the merchant in encrypted from. It is encrypted by a symmetric encryption algorithm with a randomly generated secret key K. the secret key is sent encrypted with the payment gateway's public encrypted key, $E_P$, so that only the gateway P can read it:

Customer Merchant: OI, h(PI), DS, $E_P(K)$, $E_K(P, PI, h(OI))$

The merchant forwards all elements of this message except OI to the gateway within the authorization request. He additionally includes "his" version of h (OI) so that the gateway can verify that the link between PI and OI is correct. Otherwise the customer or the merchant could claim that the payment referred to a different order than originally agreed upon. Note that this mechanism also provides a kind of payment transaction untraceability. The payment gateway can link the payment made by the same customer, nut it cannot see what was ordered. The merchant can only link the payments with order information, but cannot know which customer is behind them, provide a nonce is used as long as the payment gateway and the merchant do not conspire, dual signature provides payment transaction untraceability with respect to the merchant.

### B. Digital Signature

Digital Signatures provides information regarding the sender of an electronic document. The technology has assumed huge importance recently, with the realization that it may be the remedy to one of the major barriers to growth of electronic commerce: fear of lack of security. Digital signatures provide data integrity, thereby allowing the data to remain in the same state in which it was transmitted. The identity of the sender can also be authenticated by third parties.
The most widely used type of cryptography is public key cryptography, where the sender is assigned two keys-one public, one private, the original message is encrypted using the public key while the recipient of the message require the private key to the decrypted message. The recipient can then determine whether that data has been altered. However although this system guarantees the integrity of the message, it does not guarantee the identity of the sender (public key owner). In order to remedy this, a certificate authority is required In fig. 5.2 Juan mark (the sender) use his private key to compute the digital signature in order to compute the digital signature, a one way hashing algorithm may be used to first calculate a message digest, as is done by RSA. The message digest is an efficient way to represent a message, as well as being a unique number that can only be calculated from the contents of the massage. The sender's private key is used at this point to encrypt the massage digest. The encrypted massage digest is what is commonly called a digital signature. A certificate authority (CA) performs the task of managing key pairs, while the verification of the person or entity bound to that key pair is initially ascertained at the time of application by the registration authority.

A certificate is issued by a CA and links an individual to entity or its public key, and in some case to its private key. Certification authority can offer different grade of certificate, depending upon the type of initial identification provided by the individual.

## VI. FRESHNESS OF PAYMENT TRANSACTION MASSAGE

This service protects against replay attacked. In other worlds, it prevents eve's droppers or dishonest participants form reusing the messages exchange exchanged during a payment transaction.

### A. Nonce and time stamps

Freshness of messages can, in general, be ensured by using nonce (random numbers) and time stamps. To illustrate how they can be used in a payment transaction, here is a model based on 1KP [5] (Fig.-6.1). in the rightmost column of the figure, the names of the transaction messages are given. In 1KP there are five values that are unique for each payment transaction:

Certification Authority

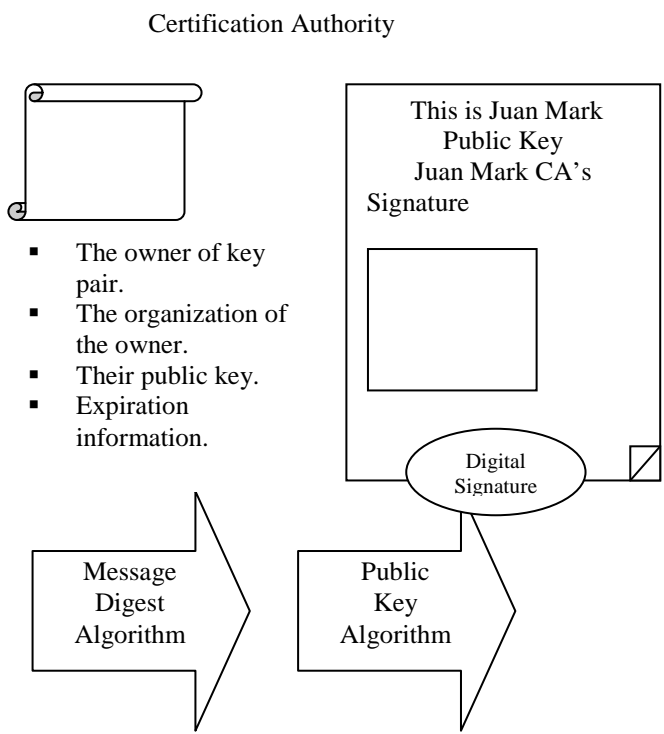

Fig. 5.2 Digital Signature Process

- Transaction identifier, $TID_M$, chosen by the merchant;
- Current date and time, DATE;
- Random number, $NONCE_M$, chosen by the merchant;
- Random number, $SALT_C$, chosen by the customer;
- Random number, $R_C$, chosen by the customer.

The purpose of TIDM, DATE, and NONCEM is to ensure freshness of all payment transaction messages except the initiate message. All three values together are referred to as TRM. All transaction messages depend on SALTC and $R_C$.

The customer initiates the payment transaction by sending the initiate message. He uses a one-time pseudonym $ID_C$
The merchant responds with the invoice message. $ID_M$ is his identifier. The value of COM represents a fingerprint of the general transaction data known by all parties:]

7COM = h (PRICE, $ID_M$, $TR_M$, $ID_C$, $h_K$ ($SALT_C$, DESC))h(.) is a collision-resistant one-way hash function, and $h_K$ (key,.) is a pseudorandom function The payment message is encrypted with the acquirer's public key $E_A$. the customer and the merchant negotiate PRICE and DESC (order information) before the initiate message. The acquirer can compute PRICE form the payment message that is forwarded to it since it encrypted with its public key $E_A$. However, it never learns DESC, since the protocol ensures confidentiality of order information with respect to the acquirer. PI is the customer's payment instruction containing, of example, this credit card number and the card's PIN.
The Auth-Request (Authorization Request) message basically contains the invoice and the payment messages. {Message} denotes the contents of the previously sent Message. The value of $h_K$ ($SALT_C$, DESC), together with COM, establishes a connection between the payment instruction and the order information. The authorization response from the acquirer and can be positive (yes) if the credit card can be charged or negative (no). the whole Auth-Response message is signed by the acquirer ($D_A$).The merchant forwards the Auth-Response message to the customer. CERTA is the acquirer's public key certificate. It can usually be retrieved online from a public directory.

## VII. AVAILABILITY AND RELIABILITY

Apart form needing to be secure, an electronic payment must be available and reliable. It must be available all the time, seven days a week, 24 hours a day. It must also have some protection against denial-of-service attacks, or at least be able to detect them early and start recovery procedures. To ensure reliability, payment transaction must be atomic. This means they occur either entirely (i.e., completely successfully) or not all, but they never hang in a unknown or inconsistent state. Furthermore, the underlying networking services as well as all software and hardware components must be sufficiently reliable. This can be implemented by adding redundancy (i.e., deliberate duplication of critical system components). Static redundancy uses n versions of a component (i.e., a function) with "m out of n voting" based on diversity. For example, with n-version programming, at least m versions must "agree" on a result to be accepted by the system as valid. With dynamic redundancy, detection of an error in one component will cause switching to a redundant component. These techniques are common to many software and hardware systems. Reliability additionally requires certain Fault tolerance mechanisms, including stable storage and resynchronization protocol for crash recovery.

## VIII. CONCLUSION

E-payment system are proliferating in banking, retail, healthcare, online markets, and even in government –in fact, anywhere money needs to change hands. Organizations are motivated by the need to deliver products and services more cost-effectively and to provide a higher quality of services to customer. Operating over the internet provides online banks with new potentialities, but also creates a set of new risks that many malicious actors are expected to use for their illegal activities. Information security, therefore, is an essential management and technical requirement for any efficient and effective payment transaction activities over the internet. This paper analyze to provide a mechanism for different transaction algorithm in e-commerce to secure the online transaction system the pay anonymity, digital signature and dual signature etc. is the best algorithm for such type of transaction the e-commerce security system is today need and demand in every where like a banking, or other business area and world wide online transactions. Now analysis of this paper gives secure transaction security for online payment system.

## REFERENCE

[1] Chaum, D., "Untraceable Electronic Mail, Return Address and Digital Pseudonyms," Comm. Of the ACM, Vol. 2, No. 24, 1981, pp. 84-88.
[2] Rubin, A. D., D. Geer, and M.J. Ranum, web Security Sourcebook. A Complete Guide to Web Security Threats and Solutions, New York, NY: John Wiley & Sons, Inc. 1997.
[3] Gulcu, C., and G. Tsudik, "Mixing E-mail with BABEL," Proc. Symp. On Network ana Distributed System Security, San Diego, CA, Feb. 22-23, 1996, pp. 2-16.
[4] E-Commerce PHI Publisher by P.T. Joseph. 2005.
[5] Fundamental of E-commerce Security by Vesna Hassler@team-fly.Artech House Computer Security Series. 2000
[6] E-Commerce Information System Series by David Whitely, Tata Mcgraw-Hill, 2006.
[7] SET Secure Electronic Transaction LLC, "The SET™ Specification," 2000.
[8] http://www.setco.org.set_specification.html.
[9] Bhattacharya, S., and R. Paul, "Accountability Issues in Multiple Message Communication,"Arizona State University, private correspondence, 1998.

AUTHORS PROFILE

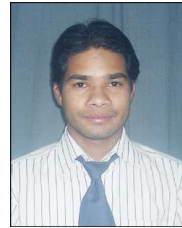

**Raju Barskar** was born in koudiya, Betul, Bhopal (M.P.), India. He is a student of M-Tech (CSE) in MANIT Bhopal. He published 2 papers in National/International Conference proceedings. His research fields include E-commerce security system, E-learning, security and optimization issues in wireless ad hoc networks, wireless sensor network, wireless mesh network and Image Processing.

**Anjana Jayant Deen** was born in Bhopal (M.P.), India. Now she is Faculty of Computer Science Department in RGPV Bhopal. Her research fields include E-commerce security system, E-learning, security and optimization issues in wireless ad hoc networks, wireless sensor network and wireless mesh network.

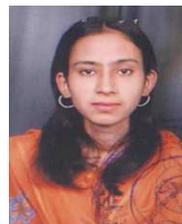

**Gulfishan Firdose Ahmed** was born in Bhopal (M.P.), India. Now she is a M.Tech (CSE) Student in the Department of Computer Science & Engineering at the LNCT Bhopal in RGPV University .She published 2 papers in National/International Conference proceedings. Her research fields includes Computer Graphics and Image Processing, E-commerce security system, Business management system, ad hoc network, wireless sensor network, computer simulation.

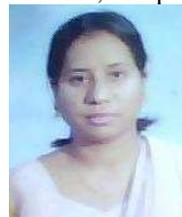

**Jyoti Bharti** was born in Bhopal (M.P.), India. Now she is an Assistant Professor in MANIT Bhopal. She published 2 papers in National/International Conference proceedings. Her research fields include E-commerce security system, Computer Graphics and image processing, wireless network and computer network.